*Granularité laser et interférences de speckles*

*Chenaud Boris* : Docteur en Physique, Agrégé de Physique
*Valvin Pierre* : Ingénieur d'Etude CNRS (G.E.S, Université Montpellier 2)

## 1. Introduction

La turbulence atmosphérique, en déformant le front des ondes lumineuses, « étale » l'image des étoiles et le pouvoir de résolution des télescopes reste alors en deçà de son maximum théorique. Il est maintenant possible de compenser ces déformations des fronts d'onde, soit en plaçant le télescope directement sur orbite hors de l'atmosphère (comme Hubble), soit en modulant la forme des miroirs du télescope pour redresser les fronts d'onde (optique adaptative). Mais il est aussi possible de jouer avec la turbulence en profitant de la cohérence spatiale de la lumière issue des étoiles. L'idée est d'utiliser le phénomène de *tavelure* encore appelé *speckle*. C'est ce que nous nous proposons d'expliquer dans ce travail. Tout d'abord nous présenterons le *phénomène de speckle et ses caractéristiques* puis nous nous pencherons sur les principes et la réalisation d'*interférences par speckles,* pour enfin aboutir à la *mesure de l'écart entre les composantes d'une étoile double* en suivant les pas de l'astrophysicien Antoine Labeyrie.

A l'époque du livre de Maurice Françon sur la granularité laser [2], ce type d'expériences nécessitait l'utilisation de plaques photographiques. Désormais, à l'heure du numérique et de la CCD, la tâche est devenue plus aisée grâce à l'utilisation des webcams. Les expériences proposées ici permettent une introduction du phénomène de speckle réalisable sur un « coin de table » et au contenu physique très riche. L'interférométrie par speckles permet d'appréhender l'*optique de Fourier* et son formalisme (produit, convolution…), sous un autre angle. Ces montages permettent également d'ancrer les notions d'*interférences à 2 ondes, à N ondes* ce qui conduit à revisiter la diffraction par les *réseaux*.

## 2. Les tavelures ou speckles

### 2.1 Présentation

Lorsqu'un laser éclaire un objet diffusant, un verre dépoli par exemple, les murs de la pièce illuminée se recouvrent alors de taches claires et sombres distribuées aléatoirement. Il s'agit de granularité laser (comme illustré sur la Figure 1). Ces « grains » de lumière sont des tavelures, des speckles, et cette figure de speckle est nette partout dans le champ et ce quelle que soit la distance d'observation. Tous les points du dépoli diffusent la lumière laser et tous ses points sont cohérents. Ces interférences aléatoires à ondes multiples forment la figure de speckle.

Les plus fins détails de cette figure de speckle, les « grains » les plus fins, ont des tailles quasiment identiques. Elles ne dépendent que des dimensions de la surface de l'objet éclairé et de l'instrument d'optique au moyen duquel on l'observe (œil, appareil photographique, lentille…) comme nous allons le voir maintenant.

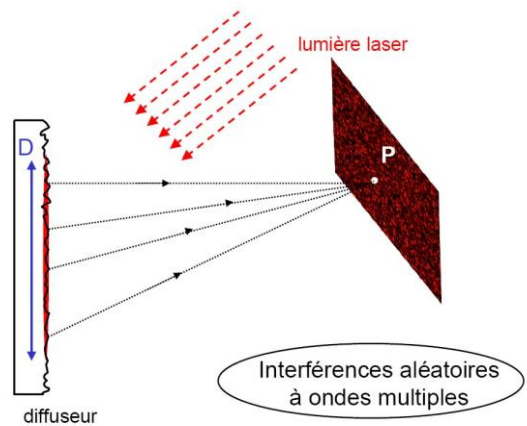

**Figure 1 : les granularités laser**

### 2.2 Taille des speckles

Nous allons ici nous intéresser au cas de la figure de speckle due à un dépoli éclairé par un faisceau laser. On observe directement les speckles sur un écran (on peut enregistrer ces speckles en remplaçant l'écran par une plaque photographique ou par un capteur CCD) sans l'intermédiaire d'une lentille ou autres.

#### 2.2.1 Approche heuristique

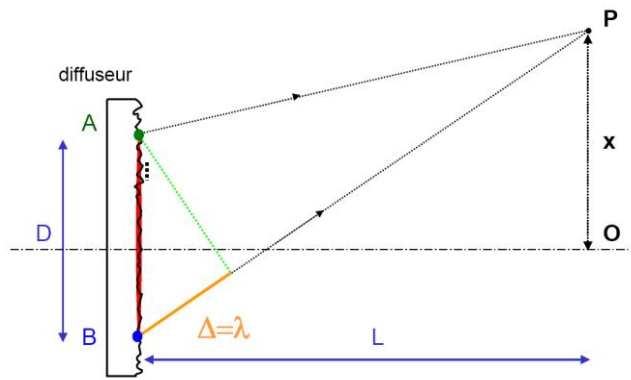

**Figure 2 : Taille des speckles**

Le laser éclaire le diffuseur sur une surface de dimension transverse notée $D$ (*cf.* Figure 2). Tous les points du dépoli de cette surface réémettent des ondes cohérentes entre elles et qui interfèrent donc. Essayons d'estimer la taille $\varepsilon$ des grains de speckle les plus fins.

Une figure de speckles est une figure d'interférences produite par les ondes (cohérentes entre elles) réémises par l'ensemble des points du dépoli contenus dans la surface éclairée. Les détails les plus fins (les speckles les plus petits) ont une dimension correspondant à « l'interfrange » le plus petit de cette figure d'interférences, noté $\varepsilon$. Ce plus fin interfrange correspond à l'interférence des ondes provenant des points de la surface éclairée les plus distants (le point A et B sur la Figure 2). En suivant les notations de la Figure 2, et en faisant l'approximation classique des petits angles ($D \ll L$ et $x \ll L$) on a au niveau du point P une différence de marche $\Delta = D.x/L$ (si O est sur la médiatrice de AB) entre les deux ondes cohérentes réémises par ces deux points. Par analogie avec l'interféromètre des trous d'Young on obtient l'interfrange $i$ des interférences à 2 ondes correspondant. On a : $i = \lambda.L/D$. Ceci nous donne l'expression suivante pour la taille $\varepsilon$ des grains les plus fins de la figure de speckle :

$$\varepsilon \approx \frac{\lambda.L}{D}$$

### 2.2.2 Montage expérimental

Le montage expérimental est celui présenté par la Figure 3. On éclaire un verre dépoli par un faisceau parallèle de lumière cohérente issue d'un laser Hélium-Néon(He-Ne). Le verre dépoli est notre milieu diffuseur. La surface « active » du dépoli est déterminée par le diaphragme réglable. Le diamètre du diaphragme est calibré, on le note *D*.

La figure de speckle est observée au moyen d'une webcam dont on a retiré l'objectif. Dans nos expériences présentées ici, nous avons utilisé une webcam *Toucam Pro 2* munie du capteur *Sony ICX098BQ*. Ce capteur est constitué de 659 x 494 pixels. Chaque pixel est un carré et le pas de notre capteur est de 5,6 µm.

L'ensemble (laser, dépoli, diaphragme et webcam) est aligné sur un banc d'optique. La distance webcam-dépoli, notée *L*, est mesurée sur ce banc. Le dépoli peut être déplacé transversalement (perpendiculairement au banc d'optique) au moyen d'une vis micrométrique. Ce dispositif nous sera utile dans la deuxième partie consacrée aux interférences.

A partir des images enregistrées on peut mesurer le diamètre des speckles[i] soit directement sur l'image, soit en utilisant une coupe « intensité-pixel » de ces images. C'est cette dernière méthode que nous avons utilisée. Les coupes « intensité-pixel » ont été obtenues à partir d'un programme écrit en langage Labview, après avoir converti ces images en tableau. La lecture d'une ligne ou d'une colonne nous donne ainsi accès à ces courbes. Pour convertir en coordonnées spatiales, il suffit de se rappeler que le pas entre chaque pixel est de 5,6 µm[ii].

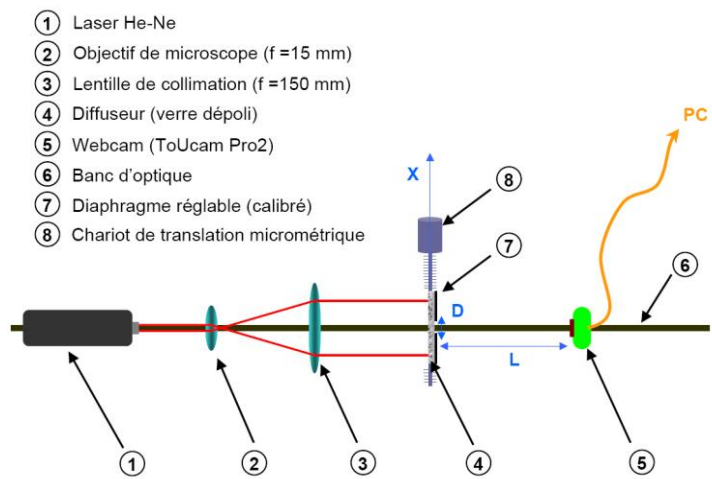

**Figure 3 : Schéma du montage expérimental**

---

[i] On mesure la taille des speckles les plus fins, ce sont eux qui contiennent l'information sur la tache éclairée: diamètre, distance, longueur d'onde du laser.

[ii] Pour le traitement des images enregistrées (courbes « intensité pixels », sommes d'images, moyennes d'images, module de la Transformée de Fourier), il est également possible d'utiliser le logiciel gratuit *Iris* développé pour l'astronomie (http://astrosurf.org/buil/us/iris/iris.htm ).

### 2.2.3 Influence du diamètre de l'ouverture

La Figure 4 présente les images obtenues pour quelques valeurs de *D*. Plus la surface « active » du dépoli est grande, plus les speckles sont fins.

Nous avons reporté sur la Figure 5 les tailles de speckles $\varepsilon$ mesurées en fonction de $1/D$. La droite correspond à la régression linéaire de $\varepsilon(1/D)$ pour la longueur d'onde $\lambda = 0,633\,\mu m$ et une distance donnée *L* = 1 m.

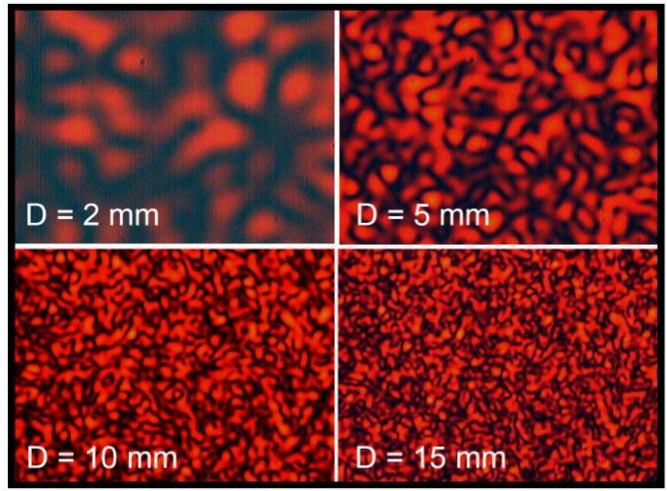

**Figure 4 : Effet de l'ouverture *D* sur la taille des speckles**

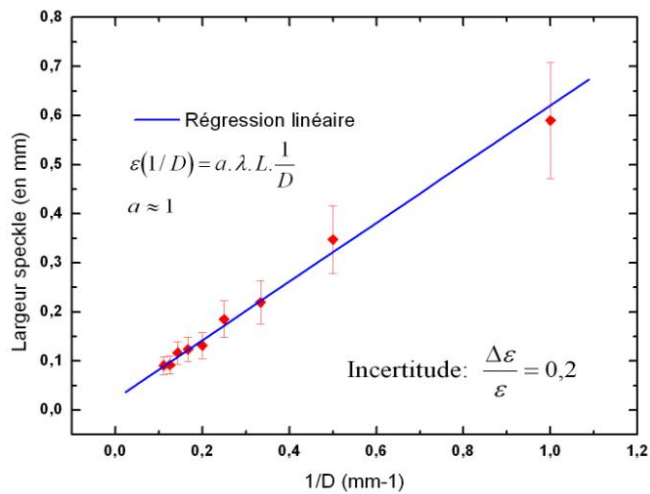

**Figure 5 : Effet de *D* sur la taille des speckles (*L* = 1 m )**

### 2.2.4 Influence de la distance d'observation

La Figure 6 présente les images de speckles obtenues pour différentes valeurs de la distance dépoli-webcam, notée *L*. La taille $\varepsilon$ des speckles est mesurée également à partir des courbes « intensité-pixels » en faisant une moyenne statistique des valeurs mesurées. Les différentes mesures de $\varepsilon$ en fonction de *L* sont tracées sur la Figure 7. La droite est la régression linéaire de $\varepsilon(L)$ pour un diamètre donné de diaphragme *D* de 1 mm et la longueur d'onde de $\lambda = 0,633\,\mu m$.

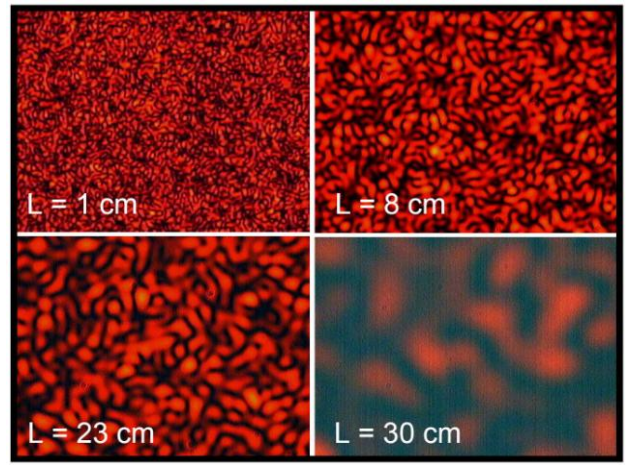

**Figure 6 : Effet de la distance d'observation sur la taille des speckles**

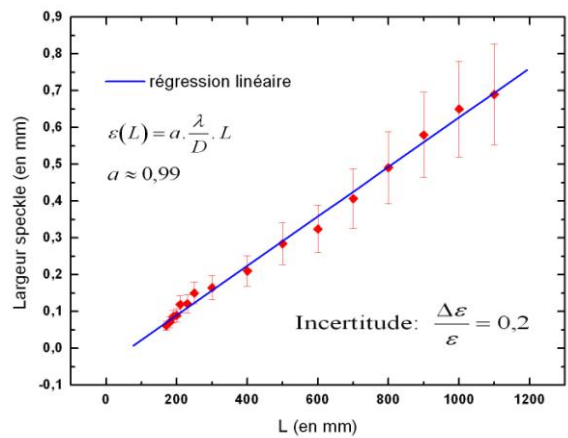

**Figure 7: Effet de *L* sur la taille des speckles (D=1mm)**

### 2.2.5 Speckles objectifs et subjectifs

Les figures de speckle que l'on a enregistrées au moyen d'une CCD ont leur dimension donnée par les caractéristiques géométriques du montage : *L*, *D*. On parle alors de *speckles objectifs* [3]. Mais lorsque l'on regarde à l'œil ces speckles, à l'intérieur de ces granularités, on observe un autre système de speckles bien plus fins et qui change lorsque l'on se déplace. On parle alors de *speckles subjectifs*. Ces speckles *subjectifs* sont dus aux interférences des taches image de tous les points de l'écran. Ils ont des dimensions caractéristiques du système optique avec lequel on les observe, en l'occurrence l'œil. Leur taille $\varepsilon'$ dépend du diamètre $D$ de la pupille de sortie du système optique et de la distance entre celle-ci et l'image du dépoli (souvent voisine de la focale image $f$). On a : $\varepsilon' \propto \lambda/D$. Pour l'œil, le diamètre de sa pupille de sortie est $D$ = 5 mm au plus (et sa focale image est d'environ f = 22 mm). La taille angulaire d'un speckle est d'environ $\alpha' \approx 30''$ (résolution angulaire de l'œil pour $\lambda = 633$ nm, s'il était optiquement parfait et pour D = 5 mm). La répartition de ces speckles subjectifs dus à l'œil se modifie lorsque l'on se déplace car l'angle sous lequel on voit l'écran et la distance œil-écran change. Ceci entraîne la modification du chevauchement des amplitudes de diffraction des différents points de l'écran et par suite celle de la figure d'interférences correspondante (*cf.* page 7 et 15 [2]). Lorsqu'on regarde à l'œil un écran recouvert par une figure de speckle ou si on enregistre l'image de cette figure de speckle au moyen d'un appareil photographique par exemple, on observe de la même façon une superposition de speckles objectifs et subjectifs.

Pour ne considérer que la figure de speckle objectif uniquement due à notre dépoli éclairé, on enregistre les figures de speckle au moyen d'une webcam dont on a retiré la lentille. On utilise ainsi directement le capteur CCD, comme on pourrait utiliser directement une pellicule avec un boîtier nu d'appareil photographique.

**3. Interférences et speckles**

### 3.1 Principe général

Dans toute cette section, on considère uniquement des figures de speckle obtenues avec un faisceau de lumière cohérente parallèle, en enregistrant les speckles par la CCD de notre webcam « nue », comme l'illustre la Figure 3.

#### *3.1.1 Translation du dépoli et figure de speckle*

En lumière parallèle, lorsqu'on translate le dépoli transversalement de $h$ selon l'axe $x$ (au moyen de la vis micrométrique), la figure de speckle enregistrée sur la CCD subit également une simple translation identique (avec $h \ll D$). Si on note $D(x',y')$ la distribution de l'intensité de la figure de speckle dans le plan de la CCD, après une translation de $h$ du dépoli selon l'axe $x$, on enregistre la distribution $D_h(x',y')$ telle que :

$$D_h(x',y') = D(x'-h, y')$$

#### *3.1.2 Superposition de speckles décalés transversalement*

On enregistre l'image d'une figure de speckle, donnant la distribution d'intensité $D(x',y')$. Ensuite le dépoli est décalé transversalement de $h$, puis on enregistre à nouveau la figure de speckle. On obtient cette fois la distribution d'intensité $D(x'-h, y')$. L'image obtenue en superposant[iii] ces deux figures de speckle, donne la distribution d'intensité $I(x',y') = D(x'-h, y') + D(x'+y')$ que l'on peut écrire (en notant $\otimes$ le produit de convolution de deux fonctions et $\delta(x',y')$ la *fonction delta de Dirac* centrée [1]) :

$$I(x',y') = D(x',y') \otimes \left[\delta(x',y') + \delta(x'-h, y')\right]$$

La *Transformée de Fourier* (notée **TF** dans la suite) $\tilde{I}(u,v)$ de cette distribution d'intensité est donc :

$$\tilde{I}(u,v) = \tilde{D}(u,v).\left[1 + e^{j2\pi.u.h}\right]$$

Ce qui peut se mettre sous la forme :

$$\tilde{I}(u,v) = 2.\tilde{D}(u,v).e^{j\pi.u.h}\left[\cos(\pi.u.h)\right]$$

Le module de cette TF s'écrit donc :

$$\left\|\tilde{I}(u,v)\right\| = 2.\left\|\tilde{D}(u,v)\right\|.\left|\cos(\pi.u.h)\right|$$

Cette image dans l'espace de Fourier correspond au module de la TF d'une figure de

---
[iii] Cette superposition des figures de speckles est réalisée soit en les sommant dans le cas d'un traitement numérique soit par une double exposition dans le cas d'un traitement analogique sur des plaques photographiques.

speckle modulée par des franges rectilignes. Cette modulation est en $|\cos(\pi.u.h)|$ ce qui donne un interfrange en $1/h$. Il s'agit d'*interférences à 2 ondes*.

Il est également possible de superposer sur une même image *N* figures de speckle décalées d'un même pas $\Delta x$. Dans ce cas, comme on le verra un peu plus loin, l'image de Fourier correspondante présente des *interférences à N ondes*. L'image de Fourier de cette superposition a les mêmes caractéristiques que la figure de diffraction de Fraunhofer d'un réseau comprenant N motifs[iv].

### 3.2 Interférences à 2 ondes

#### 3.2.1 Acquisition des images

Les pixels de notre CCD font 5,6 µm de côté. On a choisi une distance *L* relativement grande *L* > 1 m (pour se placer plutôt dans les conditions de Fraunhofer $L >> D^2/\lambda$) et une ouverture de diaphragme *D* telle que les plus fins détails des speckles soient enregistrés par au moins 5 pixels (critère d'échantillonnage). La taille des speckles avec les paramètres *L* et *D* ainsi choisis est de 30 µm. Afin de ne pas être gêné par le champ de speckles subjectifs causé par l'optique de la webcam, on effectue les différents enregistrements directement avec la CCD, l'objectif étant dévissé.

#### 3.2.2 Conversion des pixels dans l'espace réel et dans l'espace de Fourier

Le lien « pixel-espace réel » est donné par la distance entre 2 pixels : $d = 5,6$ µm. Dans l'espace de Fourier, quelle est la distance $d_{Fourier}$ entre chaque pixel ? Notre capteur comprend 659 pixels sur l'axe que l'on a choisi comme *X* et selon lequel on translate le dépoli (on peut vérifier cette valeur dans les courbes « intensité-pixels » que l'on extrait des images). La largeur totale $L_{Réel}$ de l'image réelle est donc : $L_{Réel} = 659 \times 5,6$ µm ce qui, avec le logiciel utilisé, nous donne un échantillonnage, une distance $d_{Fourier}$ entre deux pixels de l'espace de Fourier de :

$$d_{Fourier} = 1/L_{Réel} = 1/(659 \times 5,6) \text{ µm}^{-1}$$

#### 3.2.3 Mesures et TF

On enregistre une figure de speckle pour une première position du dépoli. Cette position sera dans la suite notre position de référence pour les translations du dépoli. Ensuite on décale le dépoli d'un pas *h* au moyen de la vis micrométrique (notée 8 sur la Figure 3). Puis nous avons sommé l'image décalée et l'image initiale et calculé l'image de Fourier correspondante afin de voir les interférences à 2 ondes. Les valeurs de *h* prises de manière croissante vont de 10 µm à 800 µm. Pour nos tailles de speckles, les interfranges sont bien observables à partir de $h = 100$ µm.

Nous avons calculé les sommes d'images et les TF de ces superpositions au moyen de notre programme Labview. Les images enregistrées au moyen de la CCD sont converties

---

[iv] La cohérence du laser intervient dans ce phénomène d'interférences uniquement dans la création de la figure de speckle. Les franges d'interférences rectilignes, apparaissant après le calcul de la TF, sont dues à « l'onde plane numérique » cohérente qui « éclaire » en quelque sorte notre superposition d'images (images identiques juste décalées) les images jouant un rôle analogue aux fentes d'Young. Cette « onde plane numérique » est introduite dans le calcul informatique de la TF de notre image. Pour observer physiquement ces franges rectilignes, il faudrait réaliser la figure de diffraction de Fraunhofer de cette superposition d'images (enregistrée sur une diapositive par exemple) au moyen d'une source lumineuse cohérente [2].

en tableau. On somme ensuite ces tableaux. On calcule alors le module de la TF de cette somme, puis on convertit ce dernier tableau en image 2D. Une courbe « intensité TF-pixel » peut être extraite de ce tableau. L'abscisse en pixel est finalement convertie dans l'espace de Fourier. La Figure 8 présente quelques systèmes de franges obtenus.

Dans la Figure 9, on retrouve les coupes selon l'axe horizontal de ces images. On voit très distinctement des franges. Plus le pas augmente et plus l'interfrange est serré, ce qui est rassurant. L'enveloppe de la courbe d'interférences est caractéristique de la taille des speckles. La demi-largeur de cette enveloppe est de $\Delta u = 0{,}03\,\mu m^{-1}$. Par analogie avec la *diffraction* par une *fente rectangulaire* et la *diffraction* par des grains de *poudre de lycopode* on peut retrouver la taille des speckles. Ici on obtient une largeur des plus fins speckles de $\varepsilon = 1/\Delta u = 33\,\mu m$ (autre méthode pour déterminer $\varepsilon$ par la TF de l'image du champ de speckles…). Ceci est compatible avec notre choix de L et D afin d'avoir aux alentours de 5 pixels par détail fin de notre figure de speckle.

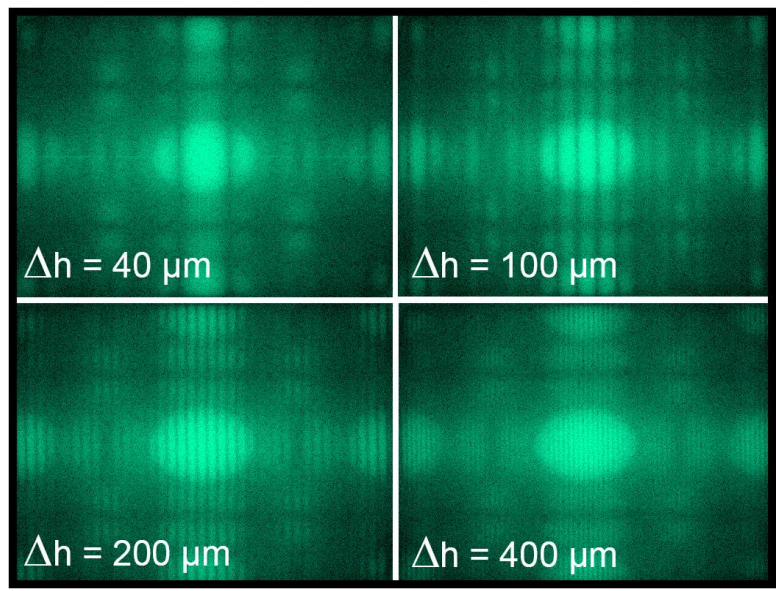

**Figure 8 : TF de la superposition de 2 figures de speckles translatées de Δx**

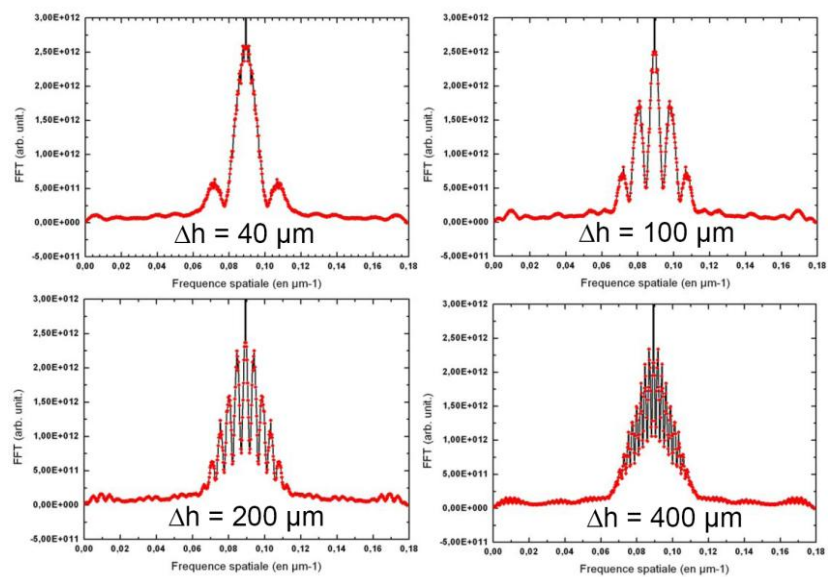

**Figure 9 : Interférences et speckles (pour 2 superpositions)**

### 3.2.4 Interférences à 2 ondes et interfranges

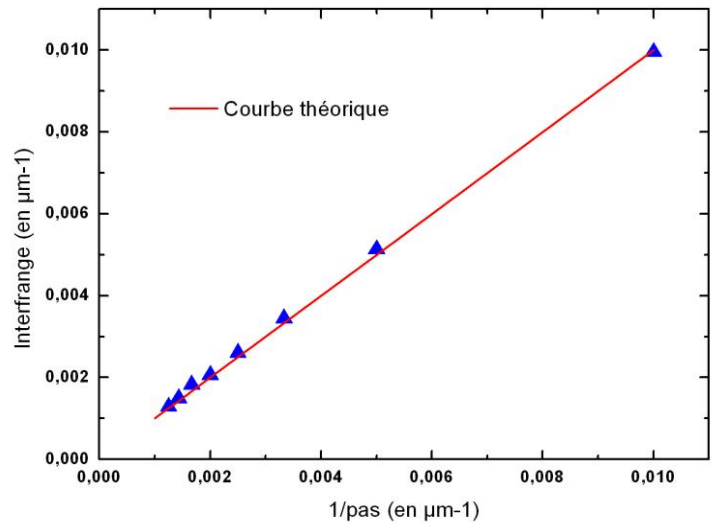

**Figure 10 : Bilan de la dépendance de l'interfrange**

La Figure 10 présente les différents interfranges mesurés en fonction de $1/h$. La droite correspond à l'expression théorique de l'interfrange $i$ (la pseudo-période) en fonction de $1/h$. On a : $i = 1/h$

## 3.3 Interférences à N ondes

### 3.3.1 Superposition de N figures de speckles et TF

Lorsqu'on superpose *N* figures de speckles décalées chacune d'un pas *h*, on obtient, dans le plan de la CCD, la distribution d'intensité $I(x',y')$. Dans le formalisme de l'optique de Fourier, on peut écrire cette distribution :

$$I(x',y') = D(x',y') \otimes \left[\delta(x',y') + \delta(x'-h,y') + \delta(x'-2h,y') + ... + \delta(x'-(N-1)h,y')\right]$$

Ou encore :

$$I(x',y') = D(x',y') \otimes \left[\sum_{k=0}^{N-1} \delta(x'-k.h,y')\right]$$

La TF de cette distribution est alors :

$$\tilde{I}(u,v) = \tilde{D}(u,v).\left[\sum_{k=0}^{N-1} \exp(2j\pi.k.h.u)\right]$$

En prenant le module de cette TF, on retrouve l'expression classique de l'étude de la diffraction de Fraunhofer par un réseau.

$$\left\|\tilde{I}(u,v)\right\| = \left\|\tilde{D}(u,v)\right\|.\left|\frac{\sin(N\pi.h.u)}{\sin(\pi.h.u)}\right|$$

Ceci n'est pas surprenant vu que, dans le formalisme de Fourier, la forme de la fonction de transmission d'un réseau est analogue à la distribution d'intensité de notre image obtenue par superposition des *N* systèmes de speckles (dans les deux expressions intervient la convolution par un peigne de Dirac de dimension *N* par le motif répété : une fente dans le cas d'un réseau, une figure de speckle dans notre cas).

### 3.3.2 Analogie avec les réseaux

La figure de ces interférences à *N* ondes est analogue à celle d'un réseau à *N* motifs de pas *h*, comme l'illustre la Figure 11.

- La période des interférences constructives reste identique. Ces maxima aux points de réception P correspondent à un déphasage de $k.2\pi$ ($k \in Z$) entre deux systèmes de speckle successifs. Dans le cas des réseaux, cela correspond à une différence de marche de $\lambda$ entre les rayons allant du point source S au point d'observation P en passant par les points homologues de deux motifs successifs.

- Plus le nombre *N* d'ondes qui interfèrent et plus le nombre *N* de superpositions de systèmes de speckles est grand, plus les pics des maxima principaux sont fins. La condition d'interférences constructives entre les différents systèmes devient draconienne au fur et à mesure que l'on augmente *N*. Par conséquent, les « régions » où l'on peut observer ces interférences constructives (les pics des maxima) sont étroites.

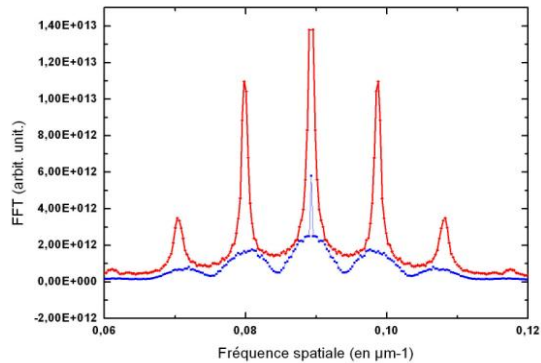

**Figure 11 : Interférences à N ondes et à 2 ondes, N = 20 et h = 100 µm**

Sur la Figure 12 sont représentées des images de Fourier pour quelques valeurs de *N*. Nous avons fait des superpositions de speckles pour des valeurs de *N* allant de 3 à 20. Pour ces différentes superpositions, le pas était invariant : $h = 100\,\mu m$.

Nous avons effectué des régressions (courbes inférieures) des coupes centrales de ces figures d'interférences. La fonction ajustée est [v] :

$$f(u) = A. \sqrt{\left[\frac{\sin(\pi.B.(u-D))}{\pi.B.(u-D)}\right]^2 . \left[\frac{\sin(N.\pi.C.(u-D))}{\sin(\pi.C.(u-D))}\right]^2}$$

où *A*, *B*, *C*, *D* sont les paramètres réglables de nos ajustements et *N* est le nombre de superpositions correspondant à chacune des courbes. Physiquement, *B* est relié à la taille de nos speckles et *C* au pas. Le premier terme, le sinus cardinal, est la TF d'un grain de speckle que l'on modélise par un carré de largeur *B*. Le second terme correspond aux interférences à *N* ondes dont nous avons calculé l'expression précédemment.

Les valeurs pour ces paramètres obtenus pour ces ajustements sont autour de 33 µm pour *B* et 106 µm pour *C*, ce qui est bien en accord avec les dimensions des grains fins de notre figure de speckle pour *B* et le décalage du dépoli pour *C*. *D* correspond à l'abscisse du centre de l'image (*i.e.* l'origine du plan de Fourier).

---

[v] La fonction $\sqrt{x^2}$ étant plus facile d'utilisation que la fonction $|x|$ équivalente avec notre logiciel de traitement (*Origine* ici).

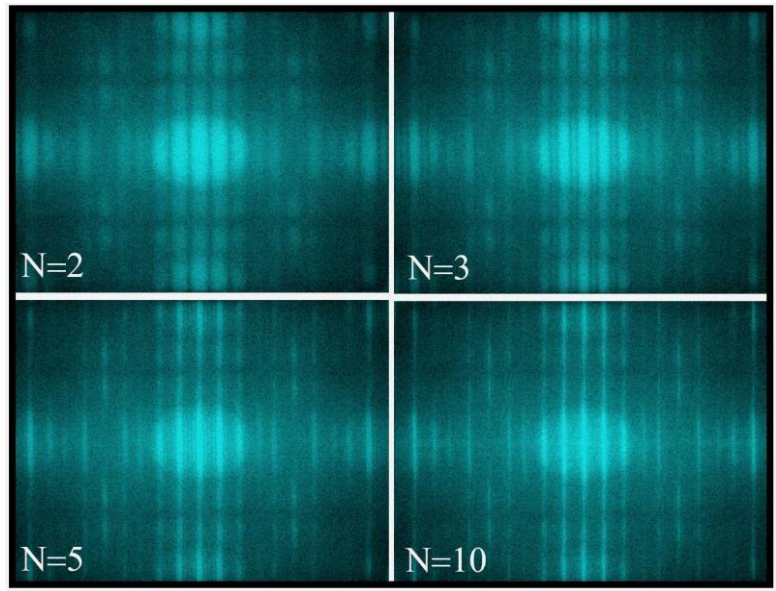

**Figure 12 : TF de *N* superpositions de systèmes de speckles corrélés**

On observe que, comme pour la figure de diffraction des réseaux, les images de Fourier de nos superpositions présentent des minima secondaires. Le nombre de ces minima secondaires est de N-2, comme le montrent les courbes de la Figure 13 et comme cela ressort sur les ajustements. On a 1 minimum secondaire pour *N* = 3, 2 minima secondaires pour *N* = 4, 3 pour *N* = 5 et 8 pour *N* = 10. Sur nos mesures, au-delà du nombre *N* = 10, on n'arrive plus à distinguer ces minima secondaires.

Si on trace la largeur des pics principaux en fonction de $1/N$, on obtient une droite comme le montre la Figure 15. On peut expliquer cette évolution par analogie avec les réseaux et les interférences à *N* ondes.

On considère un réseau comprenant un nombre $N$ pair de motifs et de pas *a*, schématisé sur la Figure 14. On observe l'intensité lumineuse au point P. Si la différence de marche $\Delta$ en P entre les ondes venant des motifs n°1 et n°*N* est $\lambda$, on a une différence de marche $\Delta' = \lambda/2$ entre celles venant des motifs n°1 et n°*N*/2. Les ondes diffractées par ces derniers motifs sont en opposition de phase en P, il y a *interférences destructives*. Il en est de même pour les ondes diffractées par les motifs n°2 et n°*N*/2 +1 et ainsi de suite. Cette position de P correspond donc à un minimum d'intensité, le premier même[vi]. Comme la différence de marche $\Delta$ varie avec $x$ comme $N.a.x/L$[vii], l'écart entre deux annulations d'intensité de part et d'autre d'un maximum secondaire est d'environ $\lambda.L/N.a$. La largeur des pics principaux décroît lorsqu'on augmente *N* en suivant une loi en $1/N$. Plus un grand nombre d'ondes interfèrent et plus la région d'accord de phase de ces ondes est étroite. C'est ce qu'on observe avec l'affinement des pics principaux.

Dans le même esprit, d'autres expériences sont possibles. Par exemple, si on déplace le diffuseur au moyen d'un moteur pendant un temps $\Delta t$ et que l'on enregistre dans un film les différentes figures de speckles correspondantes, il est alors possible de remonter à la distance de la translation et à la vitesse du mouvement. Il suffit alors de sommer les différentes images du film et de faire la TF de ce « continuum » de superpositions, ce qui

---

[vi] Le second minimum correspond à une différence de marche $\Delta = 2.\lambda$. Les ondes reçues des motif n°1 et n°*N*/4 ont alors une différence de marche $\Delta'' = \lambda/2$ et ainsi de suite…

[vii] Approximation des petits angles

nous donne un sinus cardinal (analogie avec la diffraction par une fente) dont la largeur est l'inverse du déplacement total $l$. On a alors accès à la vitesse de translation $v = l/\Delta t$.

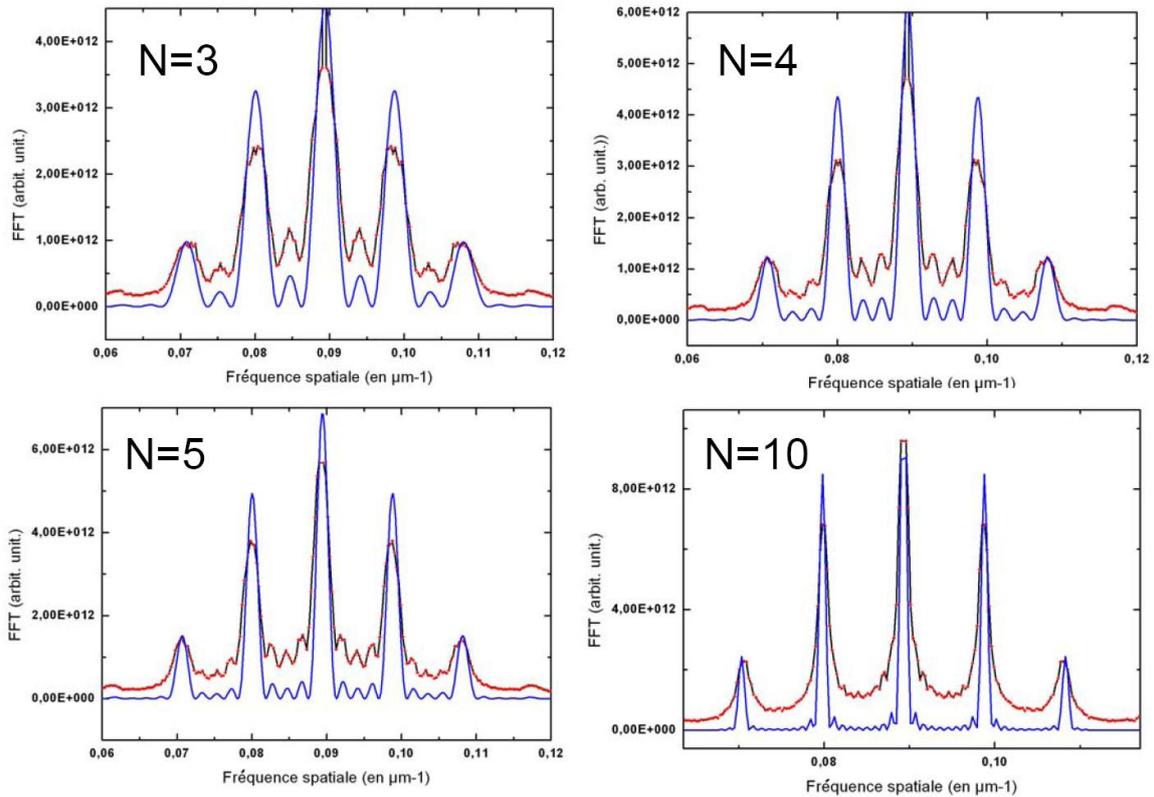

**Figure 13 : Coupe des TF et leur ajustement**

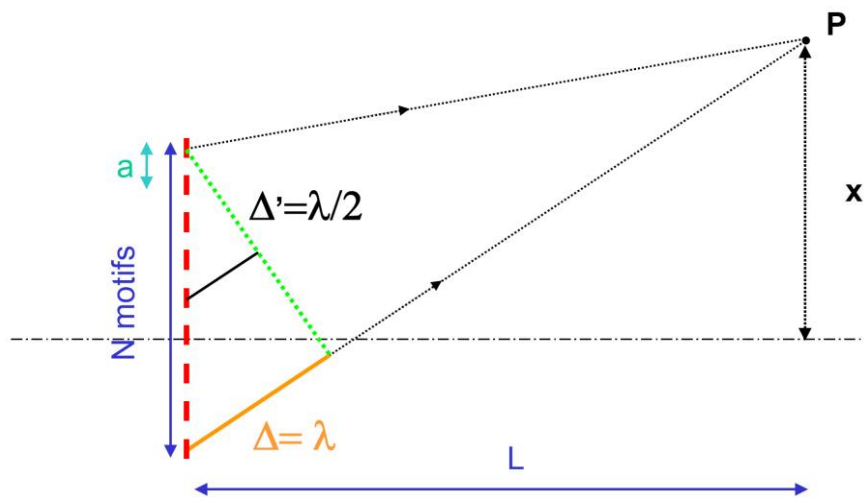

**Figure 14 : Analogie avec les réseaux à *N* motifs**

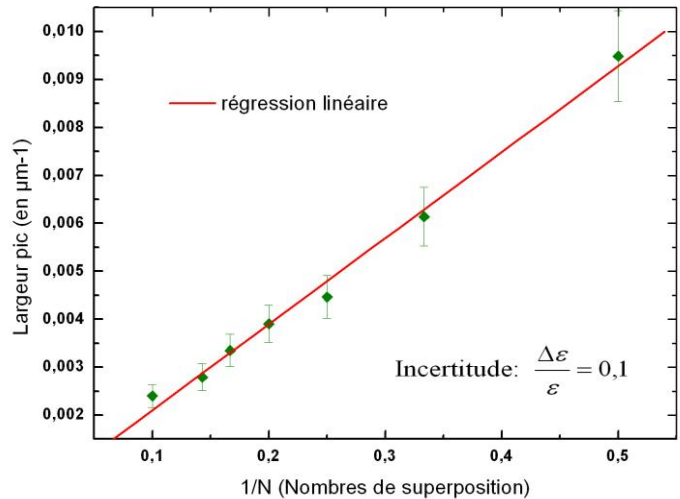

**Figure 15 : Effet du nombre *N* sur la largeur des pics**

## 4. Application : les étoiles doubles

Pour finir, nous allons présenter une application de l'interférométrie des speckles à l'astronomie. Michelson avait déjà utilisé l'interférométrie pour séparer des étoiles doubles ou encore pour mesurer leur diamètre apparent en utilisant la cohérence spatiale de la lumière reçue d'une étoile. En effet l'angle sous lequel, de la Terre, on voit toute étoile est suffisamment petit pour que l'on considère les étoiles comme des sources quasi-ponctuelles. De ce fait, la lumière de chaque étoile reçue sur Terre a une cohérence spatiale suffisante pour se prêter à des mesures interférométriques.

Une autre étape de l'interférométrie astronomique a été franchie par Antoine Labeyrie avec l'utilisation de la turbulence et des figures de speckle ainsi produites [4]. Chaque étoile produit une figure de speckle, due à la diffusion cohérente de la lumière stellaire par les turbulences de l'atmosphère. Ces turbulences atmosphériques qui dégradent fortement les observations astronomiques allaient être utilisées afin de mesurer la séparation des étoiles doubles et le diamètre des étoiles. Une nouvelle ère d'interférométrie astronomique était née, dont le VLTI (Very Large Telescope Interferometer) au Chili est un élément.

Il faut noter que l'interférométrie stellaire de Michelson et l'interférométrie des speckles de Labeyrie ne font pas apparaître les franges rectilignes à la même étape. Dans le cas de Michelson, les franges s'observent directement et ce même dans le cas d'une étoile simple. Elles sont dues à la division de l'onde issue de chacune des étoiles par deux télescopes couplés. Dans le cas de Labeyrie, on les observe indirectement par TF de l'image enregistrée et uniquement dans le cas d'une étoile double. C'est la superposition des figures de speckle corrélées de chacune des composantes de l'étoile double qui est à l'origine de ces franges dans l'image de Fourier correspondante. Dans cette dernière partie, nous présentons l'application de l'interférométrie de speckles, introduite précédemment, à la mesure de la séparation d'une étoile double, en l'occurrence le doublet *Gamma Leo (Algieba)*.

## 4.1 Montage et acquisition

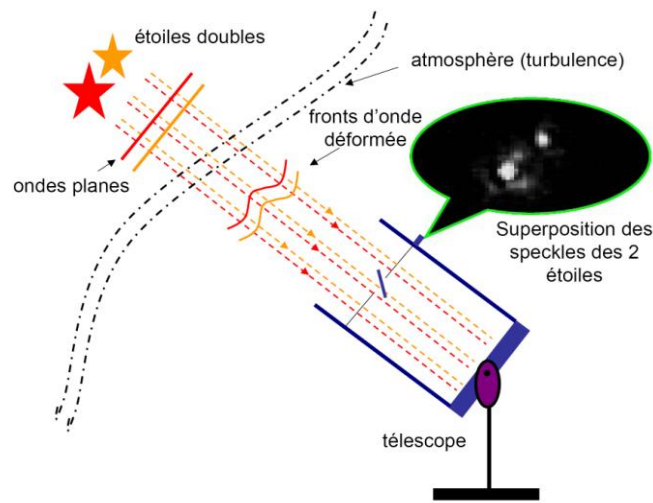

**Figure 16 : Principe de l'interférométrie de speckles « stellaires »**

Les images de ces figures de speckle ont été réalisées au moyen d'un télescope de type Makusutov-Cassegrain de 125 mm de diamètre (modèle Meade ETX 125). Le suivi a été assuré par une monture Takahashi EM200. Afin d'augmenter le grossissement, nous avons employé une lentille de Barlow x1.8. La focale résultante $f$ de notre système optique est de $f = 3420$ mm.

Notre capteur est toujours la webcam Toucam Pro 2 placée au foyer image. La largeur angulaire $\alpha$ du champ d'un pixel de ces images enregistrées est $\alpha$ = 5,6 μm / 3420 mm, ce qui donne $\alpha$ = 1,6 μrad = 0,33 ".

On ajuste le gain du capteur manuellement afin d'obtenir une image dans la dynamique de la caméra, pour un temps de pose de 1/100$^{ème}$ de seconde. Ce temps de pose permet de « figer » la turbulence.

L'acquisition des champs de speckle consiste en l'enregistrement d'un film d'une centaine d'images. Les deux étoiles formant ce doublet sont incohérentes entre elles. Chacune produit une figure de speckle qui ne peut interférer avec celle de l'autre (seules leurs intensités s'additionnent). Ces deux figures de speckle sont similaires car elles proviennent du même diffuseur, la portion d'atmosphère située entre l'étoile et le télescope. Mais elles sont décalées du fait de la séparation angulaire des 2 étoiles source. Pour mesurer cette séparation, le raisonnement précédent, pour le cas des interférences de speckles à 2 ondes, est encore applicable ici[viii]. Le module de la TF de chacune des images est donc de la forme :

$$\left\|\tilde{I}(u,v)\right\| = 2.\left\|\tilde{D}(u,v)\right\|.\left|\cos(\pi.u.h)\right|$$

C'est-à-dire que l'image de Fourier correspondante comporte des franges dont la période est $1/h$. Cet interfrange est relié à la séparation angulaire $\vartheta$ des deux étoiles.

---

[viii] Cette fois nous n'aurons besoin théoriquement que d'un enregistrement d'images puisque la superposition de speckles décalés est déjà effective comme nous avons deux sources primaires ponctuelles. Chacune de ces images comporte la superposition de la figure de speckle de chacune des deux étoiles. Une seule image est nécessaire, les figures de speckle des deux étoiles étant déjà superposées.

### 4.2 Mesures et traitement

On fait l'acquisition d'environ une centaine d'images. Au cours de cette acquisition, on souffle doucement devant le télescope de façon à augmenter artificiellement la turbulence. Les deux étoiles ne sont pas distinguables sur les images enregistrées. La Figure 17 montre la superposition de ces champs de speckles pour une acquisition longue. La turbulence n'est pas « gelée ». On observe une région lumineuse mais on ne peut rien distinguer. La caméra est orientée de telle sorte que son axe des x est selon l'axe Nord/Sud avec le Nord à gauche.

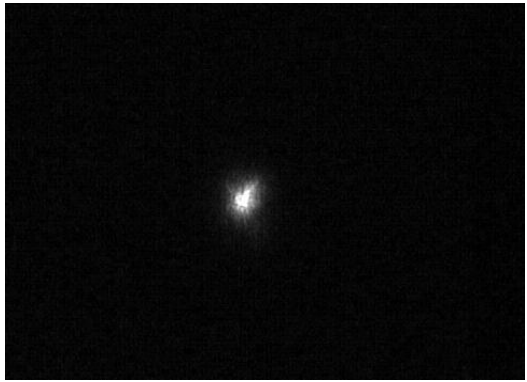

**Figure 17 : Image de l'étoile double en pause longue**

Pour ces « expériences d'interférences stellaires », nous avons traité chacune des images au moyen du logiciel IRIS. Dans un premier temps, on calcule par ce logiciel le module de la TF de chacune des 50 premières images enregistrées. On effectue ensuite une moyenne de ces 50 images de Fourier ainsi obtenues afin de lisser les effets liés aux caractéristiques stochastiques des tavelures et de mettre en évidence l'effet systématique d'apparition des franges. On obtient alors l'image présentée sur la Figure 18. Cette TF comporte bien des franges comme escompté.

#### 4.2.1 Conversion pixel-espace de Fourier

Afin de calculer les TF des images, le logiciel de traitement IRIS place l'image dans une image de 1024 x 1024 pixels. Notre image de 659 x 494 pixels est donc entourée d'un cadre noir. Par conséquent il faut un peu adapter la conversion précédente que l'on a utilisée avec Labview. Cette fois encore, puisque comme précédemment la distance entre deux pixels est de 5,6 µm, la largeur totale $L_{Réel}$ de l'image réelle est : $L_{Réel} = 1024 \times 5,6$ µm d'où la distance $d_{Fourier}$ entre deux pixels de l'espace de Fourier : $d_{Fourier} = 1/(1024 \times 5,6)$ µm$^{-1}$.

#### 4.2.2 Mesure de l'orientation du doublet stellaire

On tourne alors l'image d'un angle $\beta$ pour orienter les franges verticalement. $\beta$ est l'angle que font ces deux étoiles avec la ligne Nord/Sud. Puis en mesurant (sur l'image de Fourier tournée de $\beta$) le nombre de pixels d'un interfrange, on peut remonter à l'angle de séparation $\vartheta$ de cette étoile double.

L'angle d'inclinaison des franges que nous avons mesuré est $\beta = 124°$. Les données du catalogue WDS [5] donnent en 2003, $\beta = 126°$.

### 4.2.3 Mesure de la séparation angulaire des composantes de l'étoile double

Pour mesurer l'interfrange, on l'estime sur trois courbes « Intensité TF-pixel ». Les abscisses pour ces courbes sont celles des points de l'axe perpendiculaire aux franges (sur l'image de Fourier tournée de l'angle $\beta$, cet axe est désormais l'axe horizontal). Ces courbes « Intensité TF-pixel » sont obtenues en moyennant pour chaque point la valeur du module de la TF sur 20 pixels voisins (pris dans la direction des franges). La Figure 19 présente le type de courbes « Intensité TF-pixel » ainsi obtenues.

On mesure un interfrange de 67 pixels, avec une incertitude de l'ordre de 10 %, soit un interfrange dans l'espace de Fourier $i = 67/(1024 \times 5,6) = 0,0171 \, \mu m^{-1}$. Dans l'*espace de Fourier,* cet interfrange correspond donc dans l'*espace réel* à une « translation » de $h = 85 \, \mu m$, soit un nombre $n = 15$ de pixels.

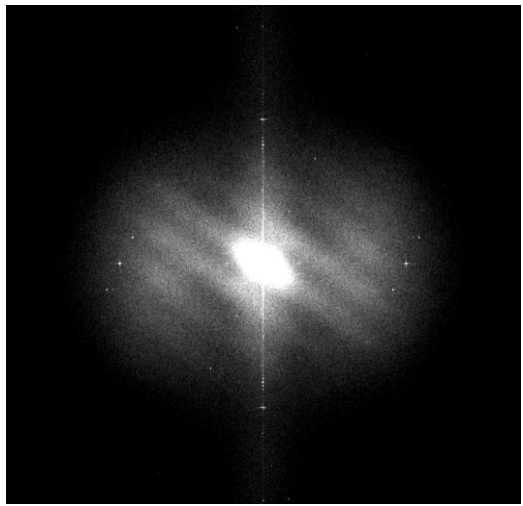

**Figure 18 : Franges d'interférences (TF des speckles de l'étoile double)**

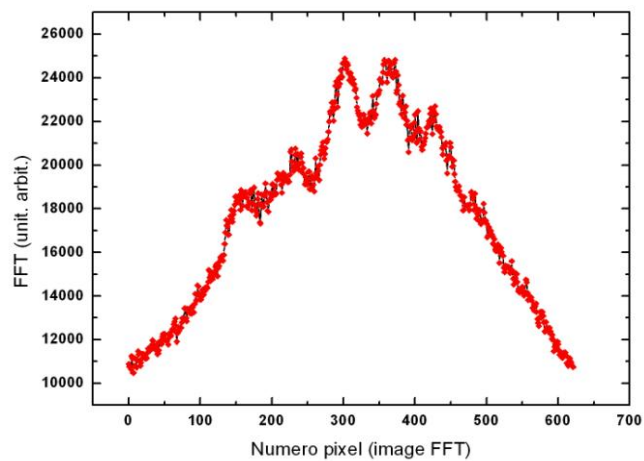

**Figure 19 : Mesure de l'interfrange (coupe de la TF )**

Par conséquent l'angle de séparation de ces étoiles doubles est de :

$$\vartheta = 15 \times 0,33 = 4,9 \text{ seconde d'arc}$$

La mesure relevée dans le catalogue WDS est de 4,5 secondes d'arc, ce qui est cohérent avec notre incertitude de 10 % sur la mesure de l'interfrange.

## 5. Conclusion

Le phénomène de speckle fut dans un premier temps un effet nuisant à l'observation astronomique, à la réalisation d'hologrammes… Mais, comme nous l'avons entrevu, la physique de ces « turbulents objets » a ouvert d'autres perspectives à l'optique : mesure de déplacements, séparations d'étoiles doubles mais aussi mesure de rugosité, encodages d'images... La popularisation des capteurs CCD via les webcams permet désormais d'envisager de nombreuses expériences sur ce thème. Les speckles amènent une approche originale de l'optique de Fourier, de l'interférométrie et raccroche ainsi, à moindre coût, l'optique à la mesure astronomique d'où elle est née.

## Bibliographie

-Optique et formalisme de Fourier :

[1] *Introduction to Fourier Optics*, J. W. Goodman, eds. Roberts & Company publishers, Denver, 2005

-Physique des speckles :

[2] *Granularité Laser (speckle) et ses applications en optique*, M. Françon, Masson, Paris, 1978

[3] *Optique cohérente (fondements et applications)*, W. Lanterborn *& al*, Masson, Paris, 2000

[4] *Attainment of Diffraction Limited Resolution in Large Telescopes by Fourier Analysing Speckle Patterns in Star Images*, A. Labeyrie, Astronomy and Astrophysics, **6**, p.85, 1970

[5] Washington Double Star catalog, B. D. Masson, G.L. Wycoff, W.I. Hartkopf, G.G. Douglass, C.E Worley, the Astronomical Journal, **122**, 3466, 2001